\documentclass[onecolumn,12pt,conference]{IEEEtran}
\normalsize

\ifCLASSINFOpdf
\else
\fi

\usepackage{longtable}
\usepackage{amsmath}
\usepackage{cancel}
\usepackage{amssymb}
\usepackage{graphicx}
\usepackage{longtable}
\usepackage{multirow}
\usepackage{array}
\usepackage[labelsep=period]{caption}
\usepackage{setspace}
\usepackage{float}
\usepackage{subfig}
\newenvironment{definition}[1][Definition]{\begin{trivlist}
		\item[\hskip \labelsep {\bfseries #1}]}{\end{trivlist}}

\newcommand{\qed}{\nobreak \ifvmode \relax \else
	\ifdim\lastskip<1.5em \hskip-\lastskip
	\hskip1.5em plus0em minus0.5em \fi \nobreak
	\vrule height0.75em width0.5em depth0.25em\fi}

\begin{document}
\title{Ergodic Capacity Analysis of Wireless Powered AF Relaying Systems over $\alpha$-$\mu$ Fading Channels}
%\author{$\textrm{Galymzhan~Nauryzbayev~and~Khaled M.~Rabie}$ \\% <-this % stops a space
%	\IEEEauthorblockA{}
	%\thanks{The authors are with the School of Electrical and Electronic Engineering, the University of Manchester, Manchester,
	%Manchester, M13 9PL UK e-mail: (see http://www.michaelshell.org/contact.html).}% <-this % stops a space
	%\thanks{Manuscript received April 19, 2014; revised January 11, 2014.}}
%}

\author{$\textrm{Galymzhan~Nauryzbayev}$,~$\textrm{Khaled~M.~Rabie}^{\ddagger}$,~$\textrm{Mohamed~Abdallah}^{*}$\\ and~$\textrm{Bamidele~Adebisi}^{\ddagger}$\\% <-this % stops a space
	\IEEEauthorblockA{$^*$Division of Information and Computing Technology, College of Science and Engineering, \\
	Hamad Bin Khalifa University, Qatar Foundation, Doha, Qatar;\\
	$^{\ddagger}$School of Engineering, Manchester Metropolitan University, Manchester, UK, M15 6BH;\\
	Email: nauryzbayevg@gmail.com;~$^*$moabdallah@hbku.edu.qa;~$^{\ddagger}$\{k.rabie;~b.adebisi\}@mmu.ac.uk}
	%\thanks{The authors are with the School of Electrical and Electronic Engineering, the University of Manchester, Manchester,
	%Manchester, M13 9PL UK e-mail: (see http://www.michaelshell.org/contact.html).}% <-this % stops a space
	%\thanks{Manuscript received April 19, 2014; revised January 11, 2014.}}
}

\maketitle

\begin{abstract}
%\boldmath
In this paper, we consider a two-hop amplify-and-forward (AF) relaying system, where the relay node is energy-constrained and harvests energy from the source node. In the literature, there are three main energy-harvesting (EH) protocols, namely, time-switching relaying (TSR), power-splitting (PS) relaying (PSR) and ideal relaying receiver (IRR). Unlike the existing studies, in this paper we consider $\alpha$-$\mu$ fading channels. In this respect, we derive accurate unified analytical expressions for the ergodic capacity for the aforementioned protocols over independent but not identically distributed (i.n.i.d) $\alpha$-$\mu$ fading channels. Three special cases of the $\alpha$-$\mu$ model, namely, Rayleigh, Nakagami-$m$ and Weibull fading channels were investigated. Our analysis is verified through numerical and simulation results. It is shown that finding the optimal value of the PS factor for the PSR protocol and the EH time fraction for the TSR protocol is a crucial step in achieving the best network performance.
\end{abstract}
%% IEEEtran.cls defaults to using nonbold math in the Abstract.


% \fi
\IEEEpeerreviewmaketitle

\section{Introduction}
\IEEEPARstart{W}{ireless} power transfer has attracted significant research interest as a potential technology to prolong the life-time of wireless battery-powered devices \cite{K1}\textendash\cite{GS2}. Exploiting radio-frequency (RF) signals for simultaneous delivery of information and power promises to be one of the main efficient techniques for wireless energy-harvesting (EH). In the literature, there are three known architectures for simultaneous wireless information and power transfer (SWIPT), namely, power-splitting (PS), time-switching (TS) and ideal relaying protocols \cite{L4}.

Over the past few years, the performance of two-hop SWIPT systems has been widely analyzed. In two-hop SWIPT systems, the relay node scavenges the energy from the received RF signal which is then utilized to forward the source information to the destination node. The authors in \cite{L4}, for instance, analyzed the performance of two-hop amplify-and-forward (AF) systems in Rayleigh fading channels. The analysis considered two relaying protocols, namely, time-switching relaying (TSR) and power-splitting relaying (PSR) protocols. In \cite{L5}, the authors evaluated the outage probability for a two-hop decode-and-forward (DF) underlay cooperative cognitive network deploying the TSR and PSR protocols, again, in Rayleigh fading channels. Furthermore, accurate analytical expressions of the ergodic capacity (EC) and achievable throughput of a DF relaying system deploying the TSR and PSR protocols over Rayleigh fading channels were derived in \cite{L6}. The outage performance in two-hop AF and DF over log-normal fading channels is studied in \cite{khaled1} and \cite{L8}, respectively. In addition, the authors in \cite{L4} and \cite{khaled2} focused on AF relaying systems with EH constraints for an ideal relaying receiver (IRR) protocol.

To the best of our knowledge, the performance of wireless EH two-hop AF relaying networks over $\alpha$-$\mu$ fading channels has not been evaluated before in the literature. Therefore, we derive new expressions for the EC in independent and not necessarily identically distributed (i.n.i.d.) $\alpha$-$\mu$ fading channels. It is worth to note that the $\alpha$-$\mu$ distribution can be regarded as a generalized model covering small-scale fading channels, namely, Rayleigh, Nakagami-$m$, Weibull, etc. The derived EC expression is unified in the sense that it considers three different relaying protocols, namely, TSR, PSR and IRR protocols. The obtained exact analytical expressions explain the behavior of these protocols under various parameters constituting various special cases of the $\alpha$-$\mu$ model, namely, Rayleigh, Nakagami-$m$ and Weibull fading channels. Our analysis is also validated by Monte Carlo simulations. Results reveal that the optimization of the EH TS and PS factors in the corresponding TSR and PSR protocols maximizes the achievable EC. Moreover, the IRR protocol achieves the best performance and the optimized PSR-based approach always outperforms the optimized TSR scheme.

The remainder of this paper is organized as follows. Section II defines the system model and its performance metric used in this paper. Sections III, IV and V derive new analytical expressions for the EC over i.n.i.d $\alpha$-$\mu$ fading channels for the TSR, PSR and IRR protocols, respectively. Section VI presents a discussion of numerical and simulation results. Finally, Section VII concludes the paper.

\section{System and Channel Model}
Consider a two-hop AF wireless communication system as shown
in Fig. 1. In this system, we assume that a source node $(S)$ transmits information to a destination node $(D)$ via energy-limited relay node $(R)$ operating in the AF mode and no direct link exists between $S$ and $D$. The relay node amplifies and then forwards the received signal to the destination. All nodes are equipped with a single antenna and operate in half-duplex mode. We assume that the relay has no external power supply (i.e., powered by energy-harvesting from source-transmitted signal). We also assume that the amount of power consumed by the relay during data processing is negligible. The source-to-relay ($S$-$to$-$R$) and relay-to-destination ($R$-$to$-$D$) links given by $h_1$ and $h_2$ are subject to quasi-static i.n.i.d. $\alpha$-$\mu$ fading. The $S$-$to$-$R$ and $R$-$to$-$D$ distances are denoted by $d_1$ and $d_2$, respectively; the corresponding path-loss exponents are given by $m_1$ and $m_2$. It is also assumed that the fading coefficients remain constant during a transmission block time $T$ but vary independently from one block to another. 

\begin{figure}[!h]
	\centering
	\includegraphics[width = 0.5\columnwidth]{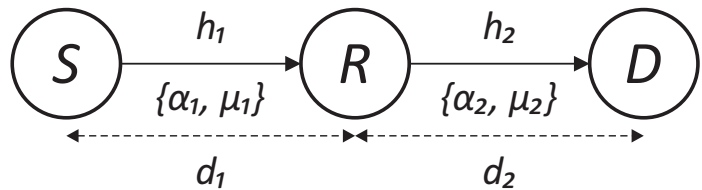}
	\caption{Two-hop AF relaying system model.}
\end{figure}

Since we assumed that the channel envelope $r$ follows the $\alpha$-$\mu$ distribution, the probability density function (PDF) of the $i-$th hop is given by \cite{galym}
\begin{equation}
\label{pdf_h_i}
f_{h_i}(r) = \frac{\alpha_i \mu_i^{\mu_i} r^{\alpha_i \mu_i - 1}}{\hat{r}^{\alpha_i\mu_i} \Gamma(\mu_i)} \exp\left( -\frac{\mu_i}{\hat{r}^{\alpha_i}} r^{\alpha_i} \right),
\end{equation}
where $\alpha_i > 0$ is an arbitrary parameter, $\hat{r}$ indicates the $\alpha-$root mean value given by $\hat{r} = \sqrt[\alpha]{\mathbb{E}\left[r^{\alpha_i}\right]}$, $\mathbb{E}\left[\cdot\right]$ stands for the expectation operator and $\Gamma\left(\cdot\right)$ denotes the Gamma function defined as $\Gamma(s) = \int_{0}^{\infty} t^{s-1} e^{-t} dt $ \cite{gradstein}. Also, $\mu_i\ge \frac{1}{2}$ is the inverse of normalized variance of $r^{\alpha_i}$ given by
\begin{equation}
\mu_i = \mathbb{E}\left[r^{\alpha_i}\right] / \left( \mathbb{E}\left[ r^{2\alpha_i} \right] - \mathbb{E}^2\left[r^{\alpha_i}\right] \right).
\end{equation}
Note that the $\alpha$-$\mu$ distribution is the most appropriate fading distribution that can be used to describe small-scale fading channels, e.g., Rayleigh ($\alpha=2$, $\mu=1$), Nakagami$-m$ ($\mu$ is the fading parameter with $\alpha=2$), Weibull ($\alpha$ is the fading parameter with $\mu=1$) \cite{magableh}.

\subsection{Ergodic Capacity}
We define the EC as
\begin{equation}
\label{erg_capacity}
\mathbb{E}\left[C_D\right] = \frac{1}{2}\mathbb{E}\left[\log_2\left(1+\gamma_D\right)\right],
\end{equation}
where $C_D$ and $\gamma_D$ indicate the capacity and signal-to-noise ratio (SNR) at the destination, respectively; the factor $\frac{1}{2}$ shows that the $S$-to-$D$ transmission requires two time slots.

\section{Time-Switching Relaying}
The principle of the TSR protocol is as follows. The time required for $S$-$to$-$D$ information transmission is given by $T$, and the time fraction designated for EH is given by $\eta T$ ($0\le\eta \le1$). The remaining time, $(1-\eta)T$, consists of two time slots to maintain the $S$-$to$-$R$ and $R$-$to$-$D$ transmissions. 

The received signal at the relay can be written as \cite{khaled1}
\begin{equation}
\label{relay_signal}
y_R(t) = \sqrt{\frac{P_S}{d_1^{m_1}}} h_1 s(t) + n_a(t),
\end{equation}
where $P_S$ stands for the source transmit power. The information signal and the noise at the relay are denoted as $s(t)$, satisfying $\mathbb{E}\left[ |s(t)|^2 \right] = 1$, and $n_a(t)$, with variance $\sigma_a^2$, respectively. Accordingly, the energy harvested at the relay can be presented as  
\begin{equation}
\label{harv_energy}
E_H^{TSR} = \theta \eta T \left( \frac{P_S}{d_1^{m_1}} h_1^2 + \sigma_a^2 \right), 
\end{equation}
where the EH conversion efficiency $\theta$ $(0<\theta\le1)$ is mainly affected by the circuitry. During the $R$-$to$-$D$ transmission, the relay performs base-band processing and then amplifies the signal which can be written as
\begin{equation}
\label{tx_relay signal}
s_R(t) = \sqrt{\frac{P_R P_S}{d_1^{m_1}}} G h_1 s(t) + \sqrt{P_R} G n_R(t),
\end{equation}
where $P_R$ represents the relay transmit power, $G$ is the relay gain defined as $G = 1/\sqrt{\frac{P_S}{d_1^{m_1}} h_1^2 + \sigma_R^2}$ and $n_R(t) = n_a(t) + n_c(t)$ denotes the overall noise at the relay with $\sigma_R^2 = \sigma_a^2 + \sigma_c^2$, where $n_c(t)$ indicates the noise added by the information receiver. Hence, the received signal at the destination can be expressed by
\begin{align}
\label{rx_dest_signal}
\hspace{-0.2cm}y_D(t) = \sqrt{\frac{P_R}{d_2^{m_2}}}G h_2 \left(\sqrt{\frac{P_S}{d_1^{m_1} }} h_1 s(t) +  n_R(t) \right) + n_D(t),
\end{align}
where $n_D(t)$ denotes the noise term at the destination with variance $\sigma_D^2$. Since the harvested energy relates to the relay transmit power as $P_R = E_H^{TSR} / \left( (1-\eta)T/2\right) $, it can be further rewritten using \eqref{harv_energy} as
\begin{equation}
\label{Pr_TSR}
P_R = \frac{2\theta \eta}{1-\eta} \left( \frac{P_S}{d_1^{m_1}} h_1^2 + \sigma_a^2 \right). 
\end{equation}

With this in mind, substituting \eqref{Pr_TSR} into \eqref{rx_dest_signal} and after some algebraic manipulations, we can write the SNR at $D$ as 
\begin{equation}
\label{dest_snr}
\gamma_D = \frac{2 \theta \eta P_S h_1^2 h_2^2}{2\theta \eta h_2^2 d_1^{m_1} \sigma_R^2 + (1-\eta) d_1^{m_1} d_2^{m_2} \sigma_D^2}.
\end{equation}

Now, to obtain an expression for EC for the TSR-based system, we first let $a_1 = 2 \theta \eta P_S,~a_2 = (1-\eta) d_1^{m_1} d_2^{m_2} \sigma_D^2,~a_3 = 2 \theta \eta d_1^{m_1} \sigma_R^2,~\mathcal{A} = a_1 X$, and $\mathcal{B} = a_2 \bar{Y},$ where $X = h_1^2$ and $\bar{Y} = h_2^{-2}$. Having these equations, the SNR $\gamma_D$ can be presented as 
\begin{equation}
\label{dest_snr1}
\gamma_D = \frac{\mathcal{A}}{\mathcal{B} + a_3}.
\end{equation}

Using \eqref{erg_capacity} and \eqref{dest_snr1}, the TSR-based EC can be written as
\begin{equation}
\mathbb{E}\left[ C_D \right] = \frac{1-\eta}{2} \mathbb{E} \left[ \log_2 \left( 1 + \frac{\mathcal{A}}{\mathcal{B} + a_3} \right)\right].
\end{equation}
The term $(1 - \eta)$ implies that the data transmission occurs only within this time fraction while the remaining time is used for EH purposes.

To simplify the EC analysis, we adopt in this paper the following lemma \cite{khairi_lemma}
	\begin{equation}
	\label{lemma}
	\mathbb{E}\left[ \ln\left( 1 + \frac{u}{v} \right) \right] = \int_{0}^{\infty} \frac{1}{s} \left( \Phi_v(s) -\Phi_{v,u}(s)\right)ds,~\forall~u,v>0,
	\end{equation}
where $\Phi_v(s)$ is the moment generating function (MGF) of the random variable (RV) $v$. $\Phi_{v,u}(s)$ can be calculated as $\Phi_{v,u} = \Phi_v(s) \Phi_u(s)$ if $v$ and $u$ are independent. 

Since $\mathcal{A}$ and $\mathcal{B}$ are independent, the EC at $D$ can be calculated using \eqref{lemma} as
\begin{equation}
\label{capacity_TSR}
	\mathbb{E}\left[ C_D \right] = \frac{1-\eta}{2\ln (2)} \int_{0}^{\infty} \frac{1}{s}\left( 1- \Phi_{\mathcal{A}}(s) \right)\Phi_{\mathcal{B}+a_3}(s)ds,
\end{equation}
where $\Phi_{\mathcal{A}}(s)$ and $\Phi_{\mathcal{B}+a_3}(s)$ indicate the MGFs of $\mathcal{A}$ and $\mathcal{B} + a_3$ given by $\Phi_{\mathcal{A}}(s) = \Phi_X(a_1 s)$ and $\Phi_{\mathcal{B}+a_3}(s) = \Phi_{\bar{Y}}(a_2 s)\exp\left( -a_3 s \right)$, respectively. Since $X$ and $\bar{Y}$ are drawn from the $\alpha$-$\mu$ distribution, we need to modify the PDF given in \eqref{pdf_h_i} to meet the changes taken on the RVs.

\begin{definition}
	Let $Z$ be a continuous RV with generic PDF $f(z)$ defined over the support $t_1 < z < t_2$ and $Q = g(Z)$ be an invertible function of $Z$ with inverse function $Z = \nu\left( Q \right)$. Then, using the \textquotedblleft change of variable\textquotedblright~method, the PDF of $Q$ can be expressed as 
	\begin{equation}
	\label{change}
	f_{Q}(q) = f_{Z}\left( \nu(q) \right)|\nu'\left(q\right)|
	\end{equation}
	defined over the support $g(t_1) < q < t(c_2)$.
\end{definition}

Since we consider $X = h_1^2$ and $\bar{Y} = h_2^{-2}$, the corresponding PDFs can be rewritten using \eqref{change} as
\begin{align}
	\label{pdf_h_1_2}
	f_{X}(r) &= \frac{\alpha_1 \mu_1^{\mu_1} r^{\frac{\alpha_1 \mu_1}{2} - 1}}{2\hat{r}^{\alpha_1\mu_1} \Gamma(\mu_1)} \exp\left( -\frac{\mu_1}{\hat{r}^{\alpha_1}} r^{\frac{\alpha_1}{2}} \right),\\
	\label{pdf_h_2_2}
	f_{\bar{Y}}(r) &= \frac{\alpha_2 \mu_2^{\mu_2} r^{-\frac{\alpha_2 \mu_2}{2} - 1}}{2\hat{r}^{\alpha_2\mu_2} \Gamma(\mu_2)} \exp\left( -\frac{\mu_2}{\hat{r}^{\alpha_2}} r^{-\frac{\alpha_2}{2}} \right).
\end{align}

The corresponding MGFs of the PDFs in \eqref{pdf_h_1_2} and \eqref{pdf_h_2_2}, which will be used later in the EC analysis for the various considered EH modes, can be calculated as 
\begin{align}
\label{MGF}
\Phi (s) = \int_{0}^{\infty}\exp\left( -s r \right) f(r) dr.
\end{align}
Substituting \eqref{pdf_h_1_2} and \eqref{pdf_h_2_2} into \eqref{MGF}, we can write the corresponding MGFs as follows
\begin{align}
\label{MGF1}
\Phi_{X} &= \frac{\alpha_1 \mu_1^{\mu_{1}}}{2\hat{r}^{\alpha_1\mu_1} \Gamma(\mu_1)} \int_{0}^{\infty} r^{\frac{\alpha_1 \mu_1}{2} - 1} e^{-s r} e^{-\frac{\mu_1}{\hat{r}^{\alpha_1}} r^{\frac{\alpha_1}{2}}} dr,\\
\label{MGF2}
\Phi_{\bar{Y}}&= \frac{\alpha_2 \mu_2^{\mu_2}}{2\hat{r}^{\alpha_2\mu_2} \Gamma(\mu_2)} \int_{0}^{\infty} r^{-\frac{\alpha_2 \mu_2}{2} - 1} e^{-s r} e^{-\frac{\mu_2}{\hat{r}^{\alpha_2}} r^{-\frac{\alpha_2}{2}}} dr.
\end{align}
These MGF integrals can be calculated in closed-form if the exponential functions in the integrands are represented in terms of Meijer's G-functions as \cite[Eq. (8.4.3.1)]{prudnikov}. Using \cite[Eq. (2.24.1.1) and (8.2.2.14)]{prudnikov}, and after some manipulations and integration steps, the MGFs of $X$ and $\bar{Y}$ are written as \eqref{mgf_h1} and \eqref{mgf_h2} at the top of the next page, where $l$ and $k$ are prime numbers such that $l/k = \alpha/2$ and $\Delta\left(\beta,\phi\right)=\left\lbrace \frac{\phi}{\beta}, \frac{\phi+1}{\beta}, \ldots, \frac{\phi+\beta-1}{\beta} \right\rbrace$. It is worth to note that the same derivation approach will be further applied for the other EH protocols. Finally, using these MGFs, we can obtain $\Phi_{\mathcal{A}}$ and $\Phi_{\mathcal{B}+a_3}$ given by \eqref{mgf_tsr1} and \eqref{mgf_tsr2}.

\begin{figure*}[!t]
	% ensure that we have normalsize text
	\normalsize
\begin{equation}
	 \label{mgf_h1}
	\Phi_{X} (s) = \frac{\alpha_1 \mu_1^{\mu_1} k^{\frac{1}{2}} l^{\frac{\alpha_1 \mu_1 - 1}{2}} s^{-\frac{\alpha_1 \mu_1}{2}}}{2 \hat{r}^{\alpha_1 \mu_1} \Gamma(\mu_1) \left( 2\pi \right)^{\frac{l+k-2}{2}}}~G_{l,k}^{k,l}\left( \left.\left( \frac{\mu_1}{\hat{r}^{\alpha_1} k} \right)^{k} \left( \frac{l}{s} \right)^{l} \right\vert \begin{array}{c}
	\Delta\left(l, 1 - \frac{\alpha_1 \mu_1}{2}\right)\\
	\Delta\left(k, 0\right)
	\end{array} \right) 
\end{equation}
	\hrulefill
	\begin{equation}
	\label{mgf_h2}
	\Phi_{\bar{Y}} (s) = \frac{\alpha_2 \mu_2^{\mu_2} k^{\frac{1}{2}} l^{-\frac{\alpha_2 \mu_2 - 1}{2}} s^{\frac{\alpha_2 \mu_2}{2}}}{2 \hat{r}^{\alpha_2 \mu_2} \Gamma(\mu_2) \left( 2\pi \right)^{\frac{l+k-2}{2}}}~G_{k+l,0}^{0,k+l}\left( \left.\left( \frac{\mu_2 k}{\hat{r}^{\alpha_2}} \right)^{k} \left( \frac{l}{s} \right)^{l} \right\vert \begin{array}{c}
	\Delta\left(l, 1 + \frac{\alpha_2 \mu_2}{2}\right), \Delta\left( k,0 \right)\\
	\textendash
	\end{array} \right)
	\end{equation}
	\hrulefill
	\begin{equation}
	\label{mgf_tsr1}
	\Phi_{\mathcal{A}} (s) = \frac{\alpha_1 \mu_1^{\mu_1} k^{\frac{1}{2}} l^{\frac{\alpha_1 \mu_1 - 1}{2}} (2 \theta \eta P_S s)^{-\frac{\alpha_1 \mu_1}{2}}}{2 \hat{r}^{\alpha_1 \mu_1} \Gamma(\mu_1) \left( 2\pi \right)^{\frac{l+k-2}{2}}}~G_{l,k}^{k,l}\left( \left.\left( \frac{\mu_1}{\hat{r}^{\alpha_1} k} \right)^{k} \left( \frac{l}{2 \theta \eta P_S s} \right)^{l} \right\vert \begin{array}{c}
	\Delta\left(l, 1 - \frac{\alpha_1 \mu_1}{2}\right)\\
	\Delta\left(k, 0\right)
	\end{array} \right) 
	\end{equation}
	\hrulefill
	\begin{multline}
	\label{mgf_tsr2}
	\Phi_{\mathcal{B} + a_3} (s) = \frac{\alpha_2 \mu_2^{\mu_2} k^{\frac{1}{2}} l^{-\frac{\alpha_2 \mu_2 - 1}{2}} ((1-\eta) d_1^{m_1} d_2^{m_2} \sigma_D^2 s)^{\frac{\alpha_2 \mu_2}{2}}}{2 \hat{r}^{\alpha_2 \mu_2} \Gamma(\mu_2) \left( 2\pi \right)^{\frac{l+k-2}{2}}}~\exp\left(-2 \theta \eta d_1^{m_1} \sigma_R^2 s\right) \times \\
	~G_{k+l,0}^{0,k+l}\left( \left.\left( \frac{\mu_2 k}{\hat{r}^{\alpha_2}} \right)^{k} \left( \frac{l}{(1-\eta) d_1^{m_1} d_2^{m_2} \sigma_D^2 s} \right)^{l} \right\vert \begin{array}{c}
	\Delta\left(l, 1 + \frac{\alpha_2 \mu_2}{2}\right), \Delta\left( k,0 \right)\\
	\textendash
	\end{array} \right)
	\end{multline}
	\hrulefill
\vspace*{4pt}
\end{figure*}

%\subsection{Outage Probability}

\section{Power-Splitting Relaying}
The PSR protocol operates over the time period, $T$, formed by two identical parts to maintain the $S$-$to$-$R$ and $R$-$to$-$D$ transmission sessions. During the $S$-$to$-$R$ transmission, the relay utilizes a portion of the received signal power for EH, $\rho P_S$, while the rest of the power, $(1-\rho)P_S$, is dedicated for the $R$-$to$-$D$ transmission. Therefore, the received signal at the input of the energy harvester can be presented as
\begin{equation}
	\sqrt{\rho}y_{R}(t) = \sqrt{\frac{\rho P_S}{d_1^{m_1}}}h_1 s(t) + \sqrt{\rho}n_a(t).
\end{equation}

The PSR-based harvested energy can be calculated as
\begin{equation}
	\label{EH_PSR}
	E_H^{PSR} = \frac{\theta \rho T}{2}\left(\frac{P_S}{d_1^{m_1}} h_1^2 + \sigma_a^2\right).
\end{equation}
This energy is used to amplify and forward the information to $D$. Thus, the relay transmit signal can be written as 
\begin{equation}
	s_R(t) = \sqrt{\frac{(1-\rho)P_S P_R}{d_1^{m_1}}} G h_1 s(t) + \sqrt{P_R} G n_R(t),
\end{equation}
where $G$ is the relay gain defined in this case as $G = 1/\sqrt{\frac{(1-\rho)P_S}{d_1^{m_1}}h_1^2 + \sigma_R^2}$ and $n_R(t) = \sqrt{1-\rho}n_a(t) + n_c(t)$. Hence, the signal at the destination can be written as
\begin{align}
\label{yD_PSR}
\hspace{-0.1cm} y_{D}(t) = \sqrt{\frac{P_R}{d_2^{m_2}}}G h_2 \left(
\sqrt{\frac{(1-\rho)P_S }{d_1^{m_1}}} h_1 s(t) 
+  n_R(t) \right) + n_D(t).
\end{align} 

The relay transmit power relates to the harvested energy as $P_R = \frac{2 E_H^{PSR}}{T}$ and therefore can be rewritten using \eqref{EH_PSR} as
\begin{equation}
\label{Pr_PSR}
P_R = \theta\rho\left(\frac{P_S}{d_1^{m_1}} h_1^2 + \sigma_a^2\right).
\end{equation}

Substituting \eqref{Pr_PSR} into \eqref{yD_PSR}, and after some algebraic manipulations, we derive the SNR at $D$ for the PSR system as
\begin{equation}
\label{snr_psr}
\gamma_D = \frac{\theta \rho (1-\rho)P_S h_1^2 h_2^2}{d_1^{m_1}\left(\theta \rho \sigma_c^2 h_2^2 + \theta \rho (1-\rho) \sigma_a^2 h_2^2 + (1-\rho) d_2^{m_2} \sigma_D^2 \right)}.
\end{equation}

Below, to obtain an expression for the EC, first we let $b_1 = \theta \rho (1-\rho) P_S,~b_2 = (1-\rho)d_1^{m_1} d_2^{m_2} \sigma_D^2,~b_3 = \theta \rho d_1^{m_1} \sigma_c^2,~b_4 = \theta \rho (1 - \rho) d_1^{m_1} \sigma_a^2,~\mathcal{K} = b_1 X$ and $\mathcal{L} = b_2 \bar{Y}$. Using these definitions, we can rewrite \eqref{snr_psr} as 
\begin{equation}
\label{snr_psr1}
\gamma_D = \frac{\mathcal{K}}{\mathcal{L} + b_3 + b_4}.
\end{equation}

Substituting \eqref{snr_psr1} into \eqref{erg_capacity}, the EC can be calculated as
\begin{equation}
\label{erg_capacity_psr}
\mathbb{E}\left[C_D\right] = \frac{1}{2}\mathbb{E}\left[\log_2\left(1+\frac{\mathcal{K}}{\mathcal{L} + b_3 + b_4}\right)\right],
\end{equation}
which, using \eqref{lemma}, can also be written as
\begin{equation}
\label{erg_capacity_psr1}
\mathbb{E}\left[C_D\right] = \frac{1}{2\ln(2)}\int_{0}^{\infty}\frac{1}{s}\left(1 - \Phi_{\mathcal{K}}(s)\right) \Phi_{\mathcal{L} + b_3 + b_4}(s) ds,
\end{equation}
where $\Phi_{\mathcal{K}}(s)$ and $\Phi_{\mathcal{L} + b_3 + b_4}(s)$ denote the MGFs of the RVs $\mathcal{K}$ and $(\mathcal{L}+b_3+b_4)$ given by $\Phi_{\mathcal{K}}(s) = \Phi_{X}(b_1 s)$ and $\Phi_{\mathcal{L} + b_3 + b_4}(s) = \Phi_{\bar{Y}}(b_2 s) \exp(-b_3 s) \exp(-b_4 s)$, respectively, and presented at the top of the next page. Now, substituting \eqref{mgf_psr1} and \eqref{mgf_psr2} into \eqref{erg_capacity_psr1} provides the PSR-based EC expression.

\begin{figure*}[!t]
	% ensure that we have normalsize text
	\normalsize
	\begin{equation}
	\label{mgf_psr1}
	\Phi_{\mathcal{K}} (s) = \frac{\alpha_1 \mu_1^{\mu_1} k^{\frac{1}{2}} l^{\frac{\alpha_1 \mu_1 - 1}{2}} (\theta \rho (1-\rho) P_S s)^{-\frac{\alpha_1 \mu_1}{2}}}{2 \hat{r}^{\alpha_1 \mu_1} \Gamma(\mu_1) \left( 2\pi \right)^{\frac{l+k-2}{2}}}~G_{l,k}^{k,l}\left( \left.\left( \frac{\mu_1}{\hat{r}^{\alpha_1} k} \right)^{k} \left( \frac{l}{\theta \rho (1-\rho) P_S s} \right)^{l} \right\vert \begin{array}{c}
	\Delta\left(l, 1 - \frac{\alpha_1 \mu_1}{2}\right)\\
	\Delta\left(k, 0\right)
	\end{array} \right) 
	\end{equation}
	\hrulefill
	\begin{multline}
	\label{mgf_psr2}
	\Phi_{\mathcal{L} + b_3 + b_4} (s) = \frac{\alpha_2 \mu_2^{\mu_2} k^{\frac{1}{2}} l^{-\frac{\alpha_2 \mu_2 - 1}{2}} ((1-\rho)d_1^{m_1} d_2^{m_2} \sigma_D^2 s)^{\frac{\alpha_2 \mu_2}{2}}}{2 \hat{r}^{\alpha_2 \mu_2} \Gamma(\mu_2) \left( 2\pi \right)^{\frac{l+k-2}{2}}}~\exp\left(-\theta \rho d_1^{m_1} \sigma_c^2 s\right) \exp\left(-\theta \rho (1 - \rho) d_1^{m_1} \sigma_a^2 s\right) \times \\
	~G_{k+l,0}^{0,k+l}\left( \left.\left( \frac{\mu_2 k}{\hat{r}^{\alpha_2}} \right)^{k} \left( \frac{l}{(1-\rho)d_1^{m_1} d_2^{m_2} \sigma_D^2 s} \right)^{l} \right\vert \begin{array}{c}
	\Delta\left(l, 1 + \frac{\alpha_2 \mu_2}{2}, \Delta\left( k,0 \right)\right)\\
	\textendash
	\end{array} \right)
	\end{multline}
	\hrulefill
	\vspace*{4pt}
\end{figure*}

\section{Ideal Relaying Receiver}
Unlike the TSR and PSR systems, the IRR-based system has the capability to independently and concurrently process the information signal and harvest energy from the same received signal. Therefore, during the first time slot, the relay harvests energy and processes the information signal whereas in the second time slot the relay uses the harvested energy to amplify and then forward the signal to the destination node. Therefore, following the same procedure shown in Section IV, we get the SNR at $D$ as
\begin{equation}
\label{snr_irr}
\gamma_D = \frac{\theta P_S h_1^2 h_2^2}{\theta d_1^{m_1} \sigma_R^2 h_2^2 + d_1^{m_1} d_2^{m_2} \sigma_D^2}.
\end{equation}

Below, to obtain an expression for the EC over $\alpha$-$\mu$ fading channels, first we let $c_1 = \theta P_s,~c_2 = d_1^{m_1} d_2^{m_2} \sigma_D^2,~c_3 = \theta d_1^{m_1} \sigma_R^2,~\mathcal{E} = c_1 X$ and $\mathcal{F} = c_2 \bar{Y}$. Therefore, we can rewrite \eqref{snr_irr} as 
\begin{equation}
\label{snr_irr1}
\gamma_D = \frac{\mathcal{E}}{\mathcal{F} + c_3}.
\end{equation}

Similar to the previous sections, the EC can be given as
\begin{equation}
\label{erg_capacity_iir}
\mathbb{E}\left[C_D\right] = \frac{1}{2}\mathbb{E}\left[\log_2\left(1 + \frac{\mathcal{E}}{\mathcal{F} + c_3}\right)\right].
\end{equation}

Also,
\begin{equation}
\label{erg_capacity_irr1}
\mathbb{E}\left[C_D\right] = \frac{1}{2\ln(2)}\int_{0}^{\infty}\frac{1}{s}\left(1 - \Phi_{\mathcal{E}}(s)\right) \Phi_{\mathcal{F} + c_3}(s) ds,
\end{equation}
where $\Phi_{\mathcal{E}}(s)$ and $\Phi_{\mathcal{F} + c_3}(s)$ denote the MGFs of the RVs $\mathcal{E}$ and $(\mathcal{F}+c_3)$ given by $\Phi_{\mathcal{E}}(s) = \Phi_{X}(c_1 s)$ and $\Phi_{\mathcal{F} + c_3}(s) = \Phi_{\bar{Y}}(c_2 s) \exp(-c_3 s)$, respectively, shown at the top of the next page. Finally, substituting \eqref{mgf_irr1} and \eqref{mgf_irr2} into \eqref{erg_capacity_irr1} provides the EC for this system.

\begin{figure*}[!t]
	\normalsize
	\begin{equation}
	\label{mgf_irr1}
	\Phi_{\mathcal{E}} (s) = \frac{\alpha_1 \mu_1^{\mu_1} k^{\frac{1}{2}} l^{\frac{\alpha_1 \mu_1 - 1}{2}} (\theta P_s s)^{-\frac{\alpha_1 \mu_1}{2}}}{2 \hat{r}^{\alpha_1 \mu_1} \Gamma(\mu_1) \left( 2\pi \right)^{\frac{l+k-2}{2}}}~G_{l,k}^{k,l}\left( \left.\left( \frac{\mu_1}{\hat{r}^{\alpha_1} k} \right)^{k} \left( \frac{l}{\theta P_s s} \right)^{l} \right\vert \begin{array}{c}
	\Delta\left(l, 1 - \frac{\alpha_1 \mu_1}{2}\right)\\
	\Delta\left(k, 0\right)
	\end{array} \right) 
	\end{equation}
	\hrulefill
	\begin{multline}
	\label{mgf_irr2}
	\Phi_{\mathcal{F} + c_3} (s) = \frac{\alpha_2 \mu_2^{\mu_2} k^{\frac{1}{2}} l^{-\frac{\alpha_2 \mu_2 - 1}{2}} (d_1^{m_1} d_2^{m_2} \sigma_D^2 s)^{\frac{\alpha_2 \mu_2}{2}}}{2 \hat{r}^{\alpha_2 \mu_2} \Gamma(\mu_2) \left( 2\pi \right)^{\frac{l+k-2}{2}}}~\exp\left(-\theta d_1^{m_1} \sigma_R^2 s\right) \times \\
	~G_{k+l,0}^{0,k+l}\left( \left.\left( \frac{\mu_2 k}{\hat{r}^{\alpha_2}} \right)^{k} \left( \frac{l}{d_1^{m_1} d_2^{m_2} \sigma_D^2 s} \right)^{l} \right\vert \begin{array}{c}
	\Delta\left(l, 1 + \frac{\alpha_2 \mu_2}{2}, \Delta\left( k,0 \right)\right)\\
	\textendash
	\end{array} \right)
	\end{multline}
	\hrulefill
\end{figure*}
\begin{figure*}[!h]
	\centering
	\subfloat[Rayleigh distribution ($\alpha=2$ and $\mu=1$).]{
		\label{subfig:Rayleigh}
		\includegraphics[width=0.3\textwidth]{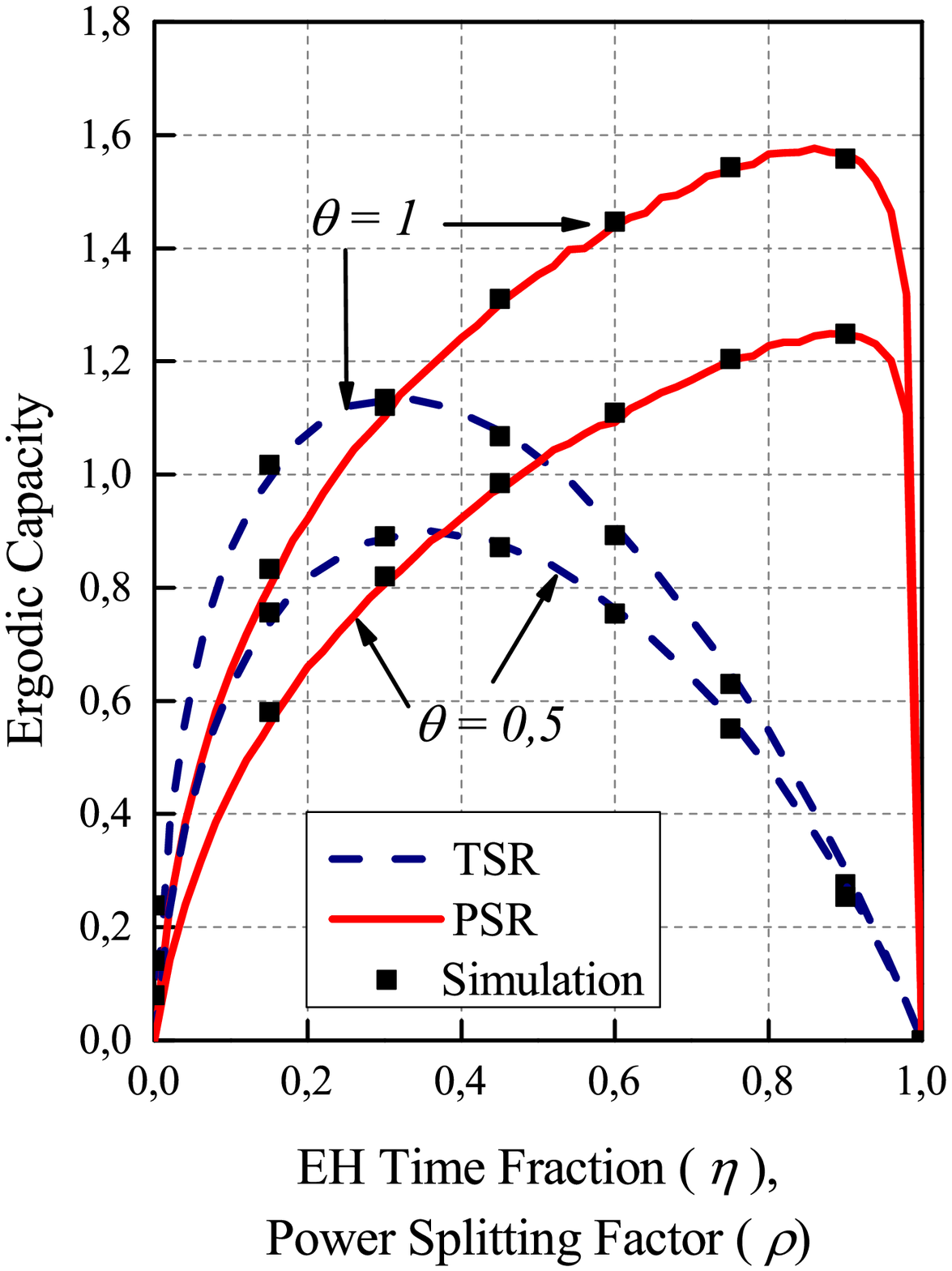}}
	\subfloat[Nakagami-$m$ distribution ($\alpha=2$ and $\mu=m=2$).]{
		\label{subfig:Nakagami}
		\includegraphics[width=0.3\textwidth]{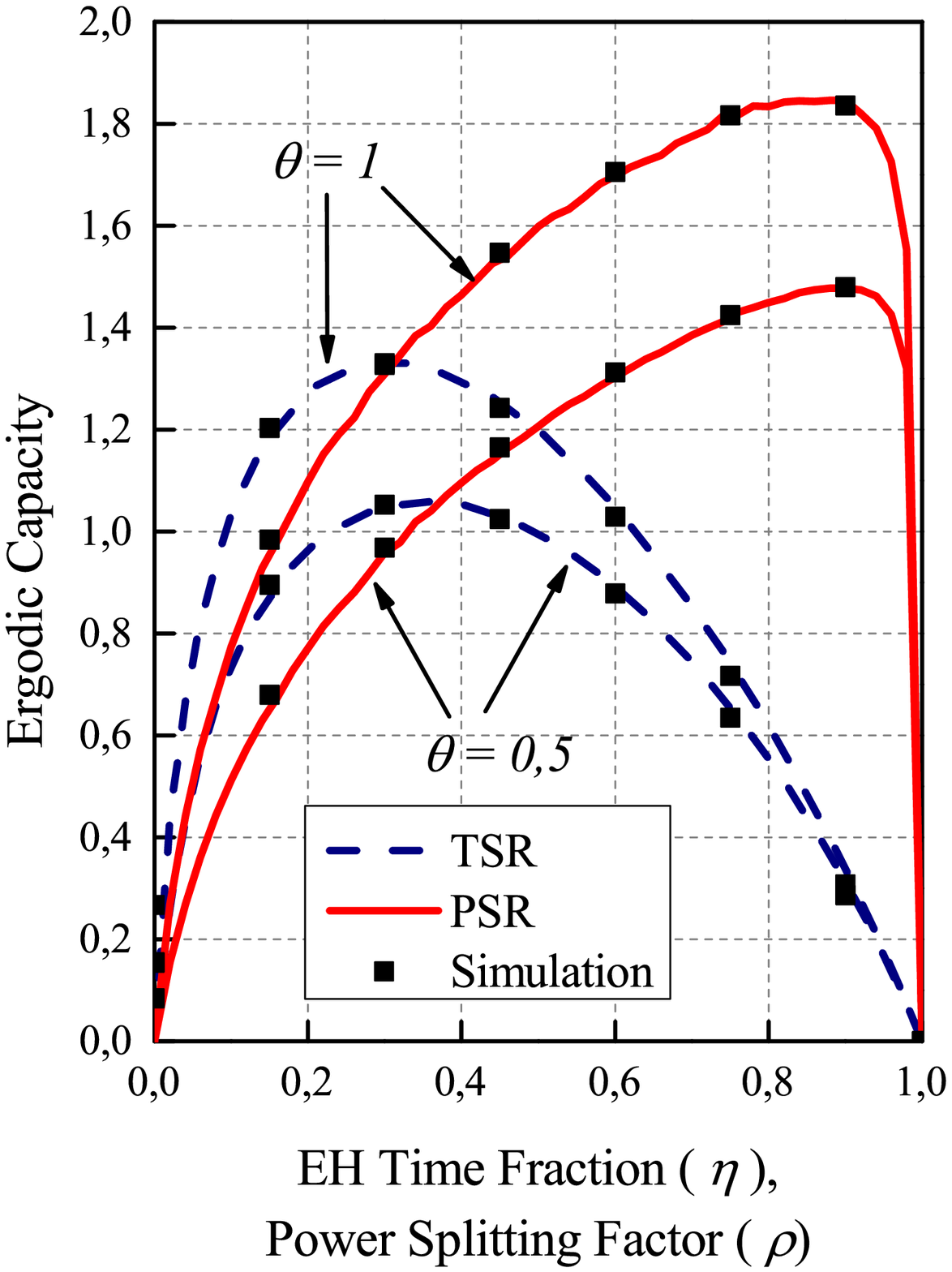}}
	\subfloat[Weibull distribution ($\alpha = 3$ and $\mu=1$).]{
		\label{subfig:Weibull}
		\includegraphics[width=0.3\textwidth]{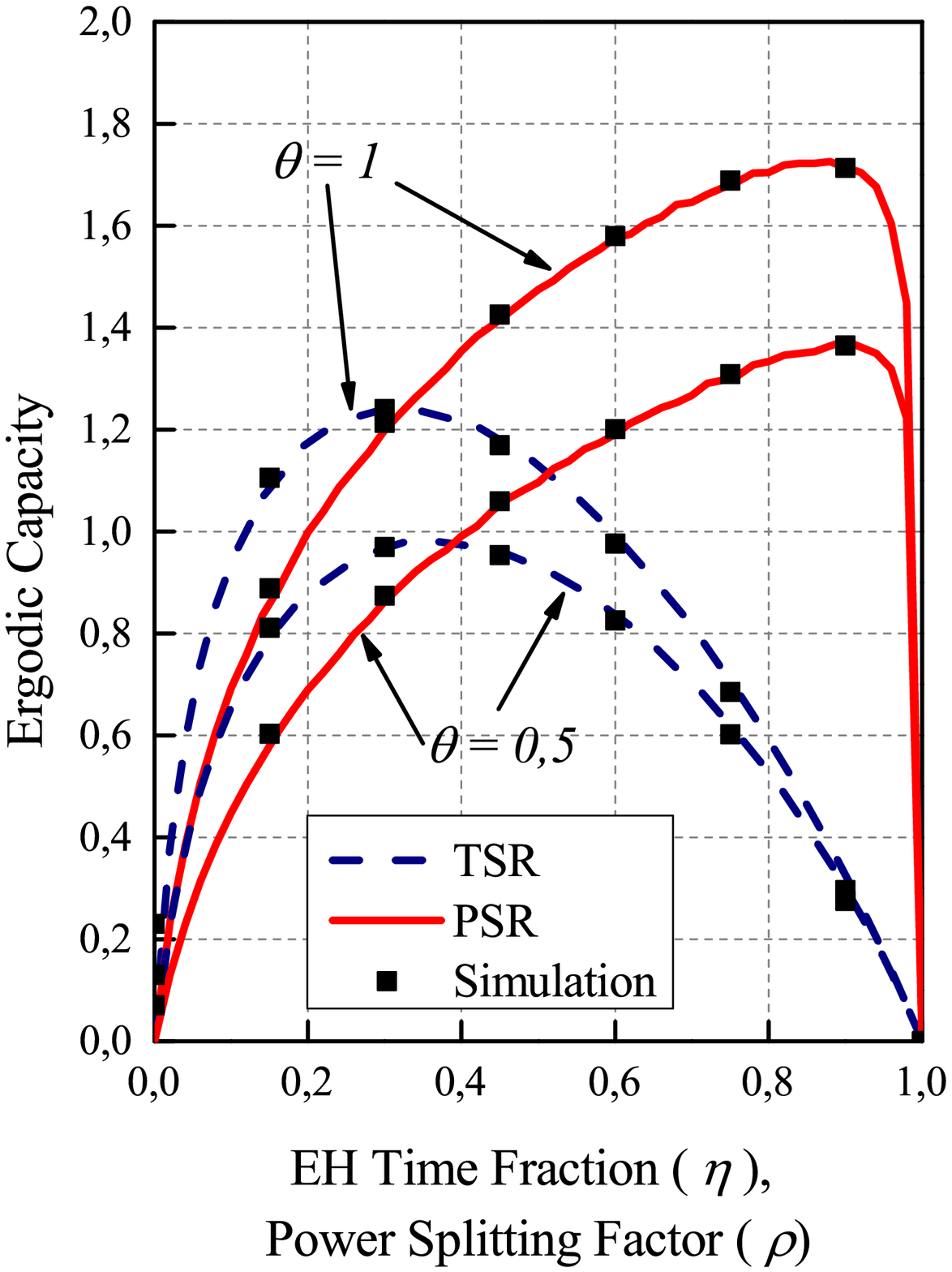}}
	\caption{EC versus the EH PS and TS factors for the PSR and TSR protocols with various parameters of the $\alpha$-$\mu$ fading channel.}
	\label{results1}
\end{figure*}

\begin{figure*}[!h]
	\centering
	\subfloat[Rayleigh distribution ($\alpha=2$ and $\mu=1$).]{
		\label{subfig:Rayleigh_d}
		\includegraphics[width=0.3\textwidth]{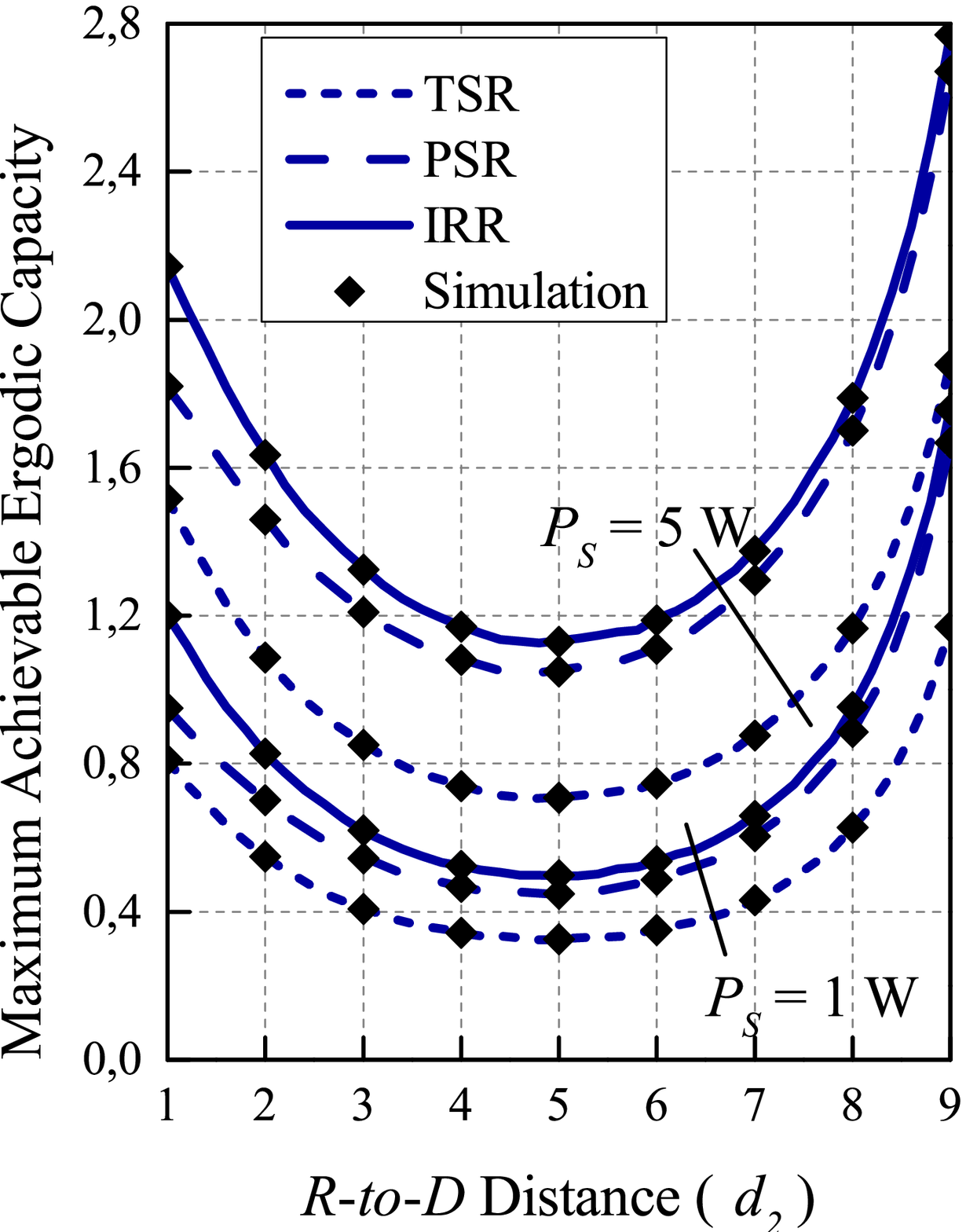}}
	\subfloat[Nakagami-$m$ distribution ($\alpha=2$ and $\mu=m=2$).]{
		\label{subfig:Nakagami_d}
		\includegraphics[width=0.3\textwidth]{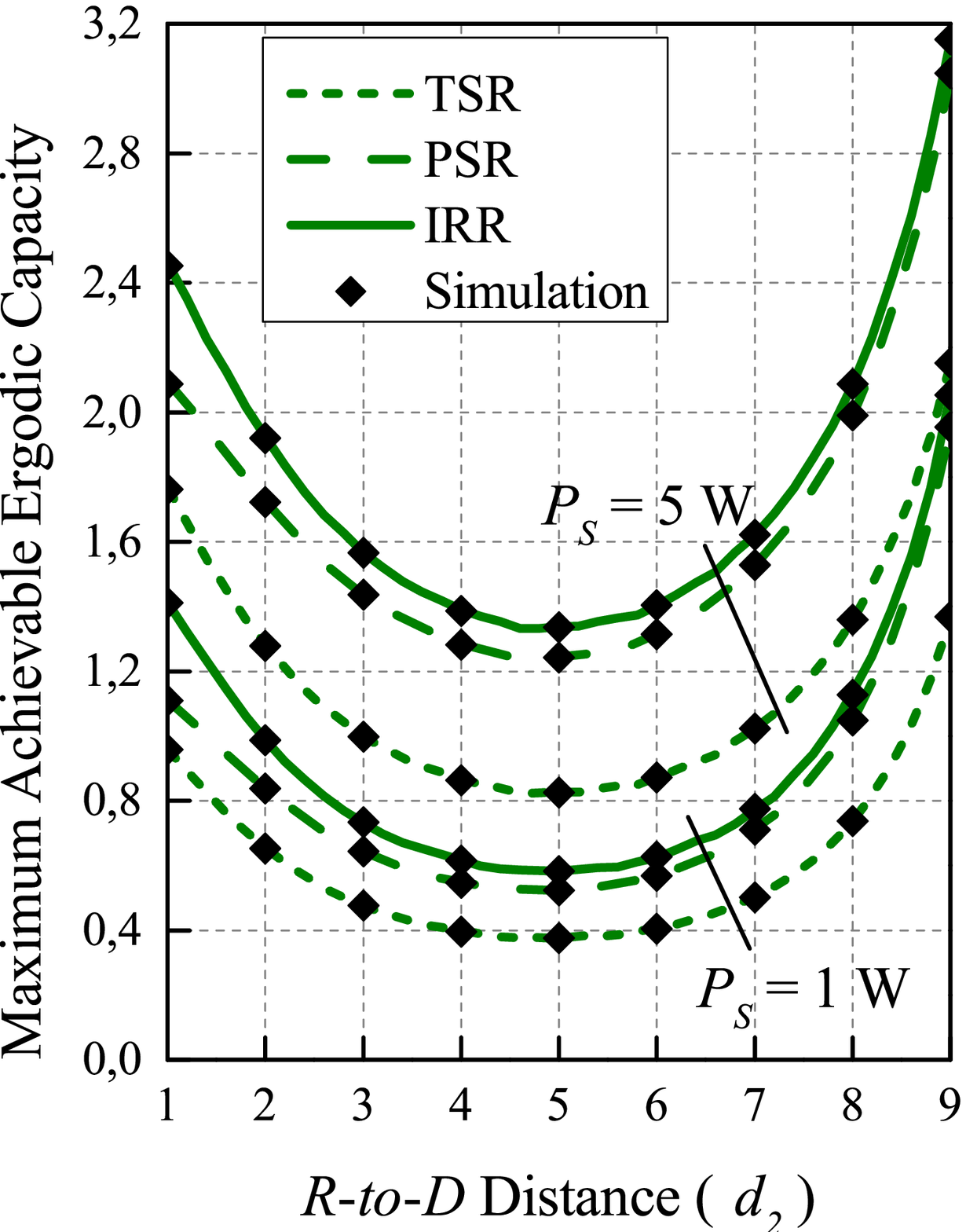}}
	\subfloat[Weibull distribution ($\alpha = 3$ and $\mu=1$).]{
		\label{subfig:Weibull_d}
		\includegraphics[width=0.3\textwidth]{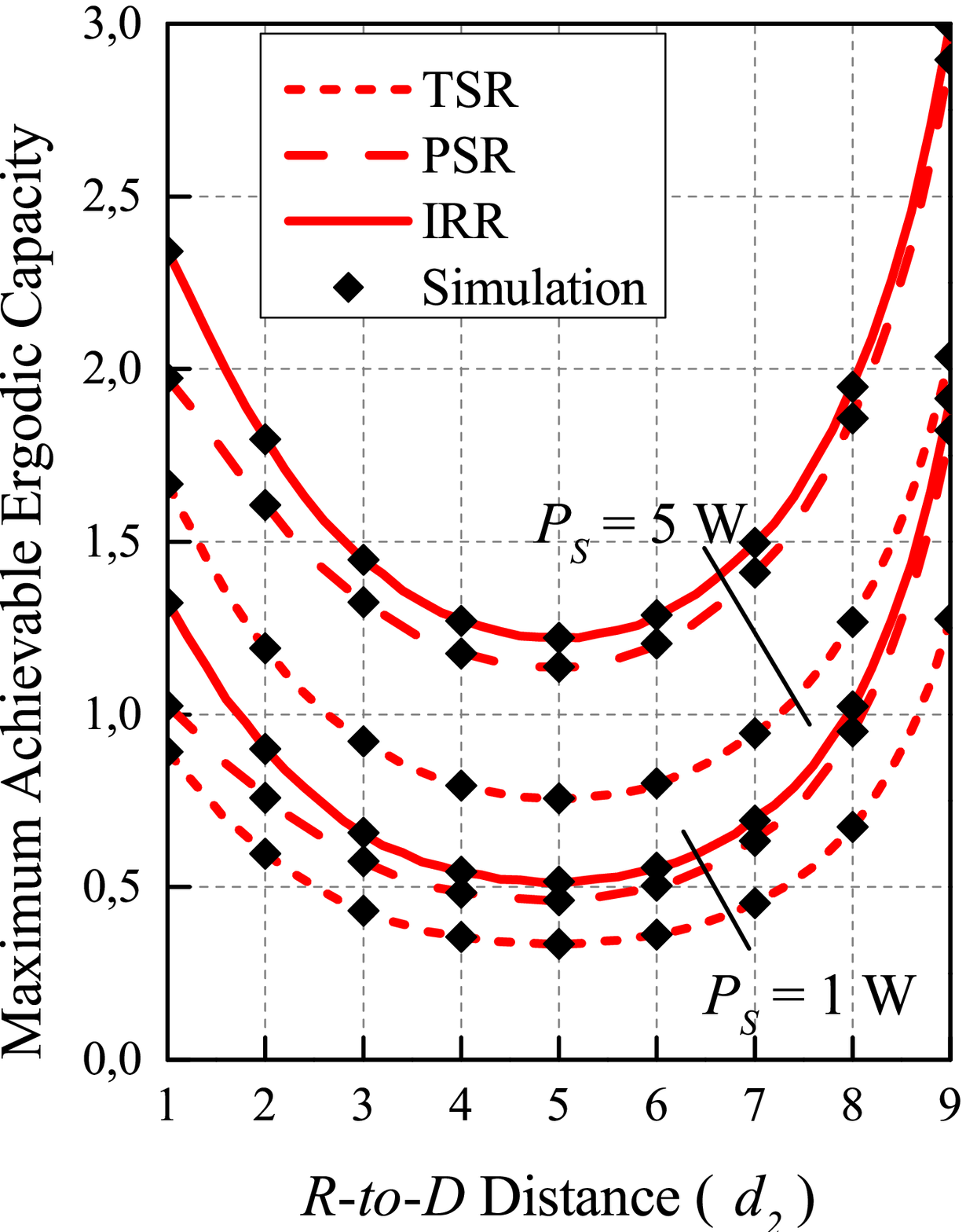}}
	\caption{EC versus the $R$-$to$-$D$ distance ($d_2 = 10 - d_1$) for the optimized TSR, optimized PSR and IRR protocols with different parameters of the $\alpha$-$\mu$ fading channel for $P_S = \{1; 5\}$ W.}
	\label{results2}
\end{figure*}
\section{Simulation Results}
This section presents numerical results for the EC expressions derived above. The adopted system parameters are as follows: $G=1$, $m_1 = m_2 = 2.7$ m, $\sigma_R = \sigma_D = 0.02$ W, $\sigma_a = \sigma_c = \sigma_R/2$. By assigning different values for the $\alpha$ and $\mu$ parameters, we derive the Rayleigh ($\alpha=2$ and $\mu=1$), Nakagami-$m$ ($\alpha=2$) and Weibull ($\mu=1$) fading coefficients.

\subsection{Ergodic Capacity}
Here, we study the impact of $\eta$ and $\rho$ on the EC for the TSR- and PSR-based systems. We consider the following network parameters: $d_1 = d_2 = 3$ m, $P_S = 1$ W and $\theta = \{0.5; 1\}$. Fig. \ref{results1} illustrates some numerical and simulation results for the ECs as a function of $\rho$ and $\eta$ for different channel distributions derived from the $\alpha$-$\mu$ distribution. Eqs. \eqref{capacity_TSR} and \eqref{erg_capacity_psr1} are used to obtain the analytical results for the TSR and PSR systems, respectively. For the case of the TSR protocol, no sufficient time is allocated for EH when $\eta$ is small and thus the relay scavenges only a small amount of power which correspondingly results in poor capacity. Being $\eta$ too large leads to the excessive amount of the harvested energy at the price of the time dedicated for information transmission which obviously results in poor capacity. The similar justification can be applied to $\rho$ for the case of the PSR protocol. It is worth to note that the selection of $\eta$ and $\rho$ mainly defines the performance of these protocols. 

\subsection{System Optimization}
Now, to assess the performance of the optimized PSR and TSR systems, first we find the optimal values of $\eta$ and $\rho$ given by $\eta^*$ and $\rho^*$ with $P_S = \{1; 5\}$ W and $\theta = 1$ for the corresponding PSR and TSR protocols by solving the following $d\left\lbrace \mathbb{E}\left[C_D\right] \right\rbrace / d\eta = 0$ and $d\left\lbrace \mathbb{E}\left[C_D\right] \right\rbrace / d\rho = 0$. It is worthwhile to mention that it is not easy to derive a closed-form solution for these equations; however, it can be calculated with software tools such as $Mathematica$.

Fig. \ref{results2} presents the maximum achievable EC according to $\eta^*$ and $\rho^*$ versus the $R$-$to$-$D$ distance when the end-to-end distance is kept fixed at 10 m. It can be seen that the optimized PSR-based system always outperforms the optimized TSR protocol irrespective of the relay location. The IRR protocol provides the best performance among all the protocols under consideration. At $d_2 = 9$ m (when the relay node is located close to the source node), the optimized PSR-based relaying approaches the performance of that of the IRR system. Moreover, the lowest EC for the three systems is detected when the relay location is midway between the source and destination nodes. This occurs due to the fact that, EH in this region achieves its peak which considerably affects the time dedicated for information transmission and hence the overall EC.

\section{Conclusion}
In this paper, we analyzed the performance of various EH relaying protocols over different i.n.i.d. $\alpha$-$\mu$ fading channels, namely, Rayleigh, Nakagami-$m$ and Weibull fadings. We derived accurate analytical expressions for the EC for the three EH systems, namely, TSR, PSR and IRR, which were validated with Monte Carlo simulations. The results showed that a proper choice of the TS and PS factors in the corresponding protocols is a key to acquire the best performance. In addition, the IRR-based system was shown to have the best performance and the optimized PSR-based approach always outperforms the optimized TSR scheme. 
\section{Acknowledgment}
This publication was made possible by NPRP grant number 9-077-2-036 from the Qatar National Research Fund (a member of Qatar Foundation). The statements made herein are solely the responsibility of the authors.

\ifCLASSOPTIONcaptionsoff
  \newpage
\fi

\end{document}